\documentclass{aa}
\usepackage{graphicx}
\usepackage{natbib}
\usepackage[figuresright]{rotating}
\usepackage{amsmath}
\usepackage{amssymb}
\usepackage{bbm}
\bibpunct{(}{)}{;}{a}{}{,}
\usepackage{txfonts}
\begin{document}
\title{Light curve analysis of Variable
stars using Fourier decomposition and Principal component analysis}

\author{Sukanta Deb\inst{1} \and  Harinder P. Singh\inst{1,2}}
\institute{Department of Physics \& Astrophysics, University of Delhi,
Delhi 110007, India
\and CRAL-Observatoire de Lyon, CNRS UMR 142, 69561 Saint-Genis Laval, France \\\email hpsingh@physics.du.ac.in
}
\offprints{H. P. Singh}
\date{}
\authorrunning{Deb \and Singh}
\titlerunning{Light curve analysis of Variable stars}

\abstract
{Ongoing and future surveys of variable stars will require new techniques for 
analysing their light curves as well as tagging objects according to their 
variability class in an automated way.
}
{We show the use of principal component analysis (PCA) and Fourier decomposition (FD) method  as tools for variable star diagnostics and compare their
relative performance in studying the changes in the light curve structures of 
pulsating Cepheids and in the classification of variable stars.
}
{We have calculated the Fourier parameters of 17,606  light curves of a variety of  variables, e.g., RR Lyraes, Cepheids, Mira Variables
and extrinsic variables for our analysis.   We have also performed PCA on the same database of light curves.
The inputs to the PCA are the  100 values of  the magnitudes for each of these 17,606 light curves in the database interpolated between  phase 0
to 1.  Unlike some previous studies, Fourier coefficients are not used as input to the PCA.}
{We show that in general, the first few principal components (PCs) are enough to reconstruct the original light curves compared to FD method where
2 to 3 times more number of parameters are required to satisfactorily reconstruct the light curves. The computation of required number of Fourier parameters on the average needs 20 times more CPU time than the computation of required number of PCs. Therefore, PCA does have some advantage over the FD method in analysing the variable stars in a larger database. However in some cases, particularly in finding the resonances in Fundamental mode (FU) Cepheids, the PCA results show no distinct advantages over the FD method. We also demonstrate that the PCA technique can be used to
classify variables into different variability classes in an automated, unsupervised way, a feature that has immense potential for larger databases
of the future.}
{
}


\maketitle
%
%
\section{Introduction}

The recent interest on the structure and properties of  light
curves of variable stars has increased a lot because of the large
flow of observational data from  variable star projects
like OGLE (Optical Gravitational Lensing Experiment), MACHO (Massive Compact Halo Object), ASAS (All Sky Automated Survey) and NSVS (Northern Sky Variability Survey). In addition, new techniques for
tagging variable objects expected in huge numbers from satellite
missions like CoRoT (Convection Rotation and Planetary Transits), Kepler, and 
Gaia in a robust and automated
manner are being explored (Debosscher et al. 2007, Sarro et al.
2009). Fourier decomposition technique is a reliable and efficient
way of describing the structure of light curves of variable stars.
Schaltenbrand \& Tammann (1971) derived UBV light curve parameters
for 323 galactic Cepheids by Fourier analysis. The first systematic
use of Fourier technique was made by Simon (1979) for analyzing the
observed light variations and radial velocity variation of AI
Velorum. The first-order amplitudes and phases from the Fourier fits
were then compared with those obtained from linear adiabatic
pulsation models to obtain the mass of AI Vel. Simon \& Lee (1981)
made the first attempt to reconstruct the light curves of Cepheid
variables using the Fourier decomposition and to describe the
Hertzsprung progression in Cepheid light curves. The method has been
applied extensively by various authors for light curve
reconstruction, mode discrimination and classification of pulsating
stars (Antonello et al. 1986, Mantegazza \& Poretti 1992, Hendry,
Tanvir \& Kanbur 1999, Poretti 2001, Ngeow et al. 2003, Moskalik \&
Poretti 2003, Jin et al. 2004, Tanvir et al. 2005). However, Fourier
decomposition by itself is not perfectly suitable for classification
of variable stars in large databases as the method works for
individual stars, but can be used as a preprocessor for other
automated schemes (Kanbur et al. 2002, Kanbur \& Mariani 2004, Sarro
et al. 2009).

The principal component analysis transforms the original data set of
variables by way of an orthogonal transformation to a new set of
 uncorrelated variables or principal components. The technique amounts to
 a straightforward rotation from the original axes to the new ones and
the principal components are derived in decreasing order of  importance
 (Singh et al. 1998). The first few
components thus account for most of the variation in the original
data (Chatfield \& Collins 1980, Murtagh \& Heck 1987). The
technique has been used for stellar spectral classification (Murtagh
\& Heck 1987, Storrie-Lombardi et al. 1994, Singh, Gulati \& Gupta
1998), QSO spectra (Francis et al. 1992) and for galaxy spectra
(Sodr\'{e} \& Cuevas 1994, Connolly et al. 1995, Lahav et al. 1996,
Folkes, Lahav \& Maddox 1996). There have been a number of studies
on the use of PCA in analyzing Cepheid light curves (Kanbur et al.
2002) and RR Lyrae light curves (Kanbur \& Mariani 2004). In both
these studies, the input data to the PCA are the Fourier
coefficients rather than the light curves themselves. Nevertheless,
it was noted that the PCA was able to reproduce the light curves
with about half the number of parameters (PCs) needed by the Fourier
technique. We may recall that in the PCA,  the first few
 PCs are usually examined as they contain most of the information
about the data. 
\begin{table*}
\centering 
\caption{Data selected for the present analysis }
\begin{tabular}{lccc}
\hline \hline 
Data&References &No. of  stars selected&Data set
\\ \hline
RR Lyrae \\ &&& \\
I band  (LMC RRab) & Soszy\'nski, I. et al. (2003) & 5835&IA\\
I band  (LMC RRc) & Soszy\'nski, I. et al. (2003)  &1751&IB\\
\hline
 Fundamental Cepheids\\  &&&  \\
I band  (LMC)  &Soszy\'nski, I. et al.(2008) &  1804&IIA\\
V band  (LMC) &Martin et al. (1979)& 6&IIB \\
V band  (LMC+SMC) & Moffett et. al (1998)&13+6&IIC \\  \hline
Overtone Cepheids \\ &&& \\
I band  (LMC) &Soszy\'nski, I. et al. (2008) &1228&III\\ \hline
Mira Variables \\&&&\\
V band&{http://archive.princeton.edu/$\sim$asas/} & 2878&IV \\\hline
Eclipsing Binary \\&&& \\
I band (LMC) &Wyrzykowski et al. (2003) & 2681 & VA\\
I band (SMC)&Wyrzykowski et al. (2004)& 1404 &VB \\
\hline \hline
\end{tabular}
\end{table*}

The  PCA  has been applied to the light curves of Cepheid variable stars by  Kanbur et al. (2002) 
and RR Lyrae stars by Kanbur \& Mariani (2004).  They concluded that PCA is more 
efficient than the FD method in bringing out changes in the light curve
structure of these variables. In our opinion, there is no advantage in the way  the PCA was  applied because the Fourier coefficients were used  as input to  
the 
PCA which are themselves the information-bearing coefficients of the  
light curve structure. Therefore PCA will not extract any  additional  information except the dimensionality reduction to a few 
orders. In the case of databases where a variety of variables are present, the method of application of PCA  on Fourier coefficients  is further complicated  
by the fact that the optimal order of 
fit to different light curves is different. When using Fourier 
coefficients as input to the PCA one has to decide where to make a cut in the
Fourier fitting orders. For Fourier 
decomposition of FU Cepheids one needs precise Fourier components 
up to order $\sim$ 10-15  in explaining the Cepheid bump 
progression whereas RR Lyraes need lesser number of Fourier components ($\sim$ 2-7 ) to completely describe the light curve structure.  
Also if the  phase coverage is not smooth then fitting of such light curves with higher order of 
the fit may give rise to wiggles and false bumps which are not associated with 
 the true light curve structures.  Therefore it is not meaningful to use Fourier coefficients as input to the PCA when 
 light curves of a large number of variable stars having different variability classes are to be analysed.
We demonstrate this fact with the following example:

 Suppose a larger database of stars contains 
RRab, RRc and FU Cepheid variables. The RRc stars are always fitted with 
lower order of the Fourier fit as compared to RRab and FU Cepheids. 
Generally RRc stars need $\sim$ 2-5  order of the fit because of  
sinusoidal and symmetric nature of their light curves,  RRab  $\sim$ 3-7  order of fit because of their asymmetric light curve whereas some of the FU
Cepheids need to be fitted with $\sim$ 10-15  order of the fit to explain the bump 
progression. Therefore for FU Cepheids, if the light curves are fitted with fewer orders,
the bump progression will not be fitted properly and one will miss 
the important bump feature. On the other hand 
if all the light curves are fitted with higher order of the fit then one is
basically fitting the noise  in the case of RR Lyrae stars which will also be 
reflected in the PCA.
 
One of the most important advantages of PCA over the FD method is that in PCA, all the light curve data can be processed and analysed 
 in one go if all the phased light curve data can be made of similar dimensions as we shall demonstrate later, 
 whereas in the FD method each light curve has to be fitted with optimal order of the fit and analysed
individually.  This is a very time consuming process for large databases. Therefore, the 
decision regarding the cut in the order of the fit is manual and hence very 
cumbersome.  Unlike FD,  one can decide where to make a 
cut in the PCA order in light curve reconstruction for all the light curves  
simply by looking at the cumulative percentages of variance in the data set.
The optimal data compression using PCA is enormous,  a fact that
is quite relevant with the larger databases of the future.

PCA has also the advantage of preferential removal of noise from 
the light curve data and isolating the bogus light curves,  
whereas for precise Fourier decomposition, one needs very 
well-defined and accurate light curves free from noisy, scattered data  
points and having a good phase coverage. The most significant PCs contain those features which are most strongly 
correlated in many of the light curves. Therefore, the noise which is 
uncorrelated with any other features will be represented in the less 
significant components. Also by retaining only the most significant PCs 
to represent the light curves we achieve a data compression that preferentially
removes the noise.  PCA can be used to filter out 
bogus features in the data as it is sensitive to the relative frequency of 
occurrence of features in the data set ( Bailer-Jones et al. 1998).
However, one distinct disadvantage of PCA is that addition of a single light curve in the analysis requires the entire PCA to be redone.

In this paper, we show the use of PCA directly
on the light curve data of more than 17,000 
stars (RR Lyraes , Cepheids, Eclipsing binaries and Mira variables) taken from 
the literature and different existing databases. We also apply the FD
method to these light curves to determine the Fourier 
parameters. Denoising should be carried out before the Fourier decomposition if the light curves are noisy. However, the photometric error  
in the light curves in the case of the present selected database is very small, 
i.e., the light curves  data have a good photometric accuracy( $\sim$  
0.006 - 0.14 mag in the case of OGLE database and $\sim$ 0.02 - 0.220 mag in the case of  ASAS database ). To investigate the noise in the light  
curves we have calculated the unit-lag
auto-correlation function on the residual light curves. The auto-correlations
are found to be $<<$ 1. Therefore no denoising has been carried out. However, in
some light curves there were outliers present. To remove these outliers,
we have used a robust multi-pass non linear fitting algorithm in IDL (Interactive Data Language). We use light curves (magnitudes at different epochs) 
as input to PCA  and compare relative performance of the ability of PCA in 
finding resonances in Cepheids and in the classification of different types of 
variables as compared to the FD method. 
We have, therefore, performed independent automated Fourier analysis of all the data sets described in the paper  using a computer code developed by us.

Another aim of this paper is  to analyze the 
performance of PCA as a fast, automated and unsupervised classification tool 
for variable stars. Since one of the important aspects of this paper is to do a preliminary PCA based classification in an unsupervised way  on a 
larger sets of astronomical data, we explore the possibility of its use for future databases. 
PCA can be used for preliminary classification of the variable stars such as 
classification between  pulsating stars  and Eclipsing binaries and different 
variability classes.

We present the Fourier decomposition  technique using
Levenberg-Marquardt algorithm for non-linear least square fitting
(Press et al. 1992) in Sect. 2. We also describe the unit-lag
auto-correlation function for finding the optimal order of the
fit. Sect. 3 describes the PCA for dimensionality reduction and light curve
 reconstruction. Sect. 4 describes the results obtained by the FD and PCA 
techniques when applied to study the structure of Cepheid light curves. In addition, we compare the relative performance of FD and PCA for classification of various variability classes in the database selected for the 
present analysis. Finally in Sect. 5, we present
important conclusions of the study.

\section{Fourier Decomposition technique}

Since the light curves of the selected ensemble of  variable stars are periodic, they can
be written as a sum of cosine and sine series:
\begin{equation}
 m(t) =A_{0}+ \sum_{i=1}^{N}  a_{i}\cos(i \omega (t-t_{0}))+ \sum_{i=1}^{N} b_{i}\sin(i \omega (t-t_{0})),
\end{equation}
where $m(t)$ is the observed magnitude at time $t$, $A_{0}$ is the
mean magnitude, $a_{i}$, $b_{i}$ are the amplitude components of
$(i-1)^{th}$ harmonic, $P$ is the period of the star in days, $\omega$=2$\pi/P$ is the angular frequency, 
and $N$ is the order of the fit. $t_{0}$ is the epoch of maximum light.
Obviously, Eq.~(1) has $2N +1$ unknown parameters which require at least 
the same number of data points to solve for these parameters. Equivalently, 
we can write Eq.~(1) as
\begin{equation}
 m(t) =A_{0}+ \sum_{i=1}^{N}  A_{i}\cos(i \omega (t-t_{0})+\phi_{i}),
\end{equation}
where $A_{i} = \sqrt{{a_{i}}^2+{b_{i}}^2}$  and $\tan \phi_i = -
b_i/a_i$. Since period is known from the respective databases, the observation time can be folded
into phase ($\Phi$) as (cf. Ngeow et al. 2003)
\begin{equation}
\Phi =\frac{\left( t-t_{0}\right) }{P}-Int\left( \frac{ t-t_{0}
}{P}\right),
\end{equation}
The value of $\Phi$ is
from 0 to 1, corresponding to a full cycle of pulsation and $Int$ denotes the integer part of the quantity. 
Hence, Eqs.~ (1) and (2) can be written as (Schaltenbrand \& Tammann
1971)
\begin{equation}
 m(t) =A_{0}+ \sum_{i=1}^{N}  a_{i}\cos(2\pi i \Phi(t))+ \sum_{i=1}^{N} b_{i}\sin(2\pi
  i \Phi(t)),
\end{equation}
\begin{equation}
m(t) = A_{0}+\sum_{i=1}^{N}  A_{i} cos[2\pi i \Phi(t)+\phi_{i}],
\end{equation}
with relative Fourier parameters as
\[ R_{i1}=\frac{A_{i}}{A_{1}} \\; \, \phi_{i1}=\phi_{i}-i\phi_{1}\]
where $i > 1$. The combination of coefficients $R_{i1}$, $\phi_{i1}$
where $i=2, 3, 4...$ can be used to describe the progression of
light curve shape in the case of Cepheids and  other variables
and can be used for variable star classification. In Table 1, we list
all the variable star light curve data that has been subjected to the analysis.
In the case of the data taken from the OGLE database (Soszy\'{n}ski et al. (2003, 2008) and Wyrzykowski et al. (2003, 2004), the number of stars seems 
to be more than the actual number presented in the database. This is because 
of the fact that we have not tried to remove the overlapping stars in  
different OGLE fields as this will not affect our analysis. In the case of  data from Martin et al. (1979) , the stars 
with poor phase coverage have been left out.

The estimation of optimal number of terms to be used in the Fourier
decomposition of  the individual light curve is not straightforward. 
As has been pointed out by Petersen (1986), if $N$ is
chosen too small, a larger number of Fourier parameters can be
calculated from a given observation and the resulting parameters
will have systematic deviations from the best estimate. On the other
hand, if N is chosen too large, we are fitting the noise. Following
Baart (1982), Petersen (1986) adopted the calculation of unit-lag
auto-correlation of the sequence of the residuals in order to decide
the right $N$ so that the residuals consist of noise only. It as
defined as
\[ \it \rho := \frac {\sum_{j=1}^{n} \left(v_{j}-\bar{v}\right)
\left(v_{j+1}-\bar{v}\right)}{\sum_{j=1}^{n} {(v_{j}-\bar{v})}^2}
\]
 where $v_{j}$ is the $j^{th}$ residual,
$\bar{\it v}$ is the average of the residuals and $j=1,....n$ are
the number of data points of a light curve. The value of {\it v} is
basically the residuals of the fitted light curve
\[v=m(t)-[A_{0}+\sum_{i=1}^{N}  A_{i} cos(2\pi i \Phi(t)\ +
\phi_{i})]\]

 It should be noted that for the calculation of $\rho$
we must choose the ordering of $v_{j}$ given by increasing phase
values rather than ordering given by the original sequence. A
definite trend in the residuals will result in a value of ${\it
\rho}$ equal to 1, while uncorrelated residuals give smaller values
of ${\it \rho}$. In the idealized case of residuals of equal
magnitude with alternating sign, ${\it \rho}$ will be approximately
equal to $-1$. The suitable value of ${\it \rho}$ can be chosen
using Baart's condition. According to this, a value of ${\it \rho}
\geq$ $[n-1]^{-1/2}$ (where $n$ is the number of observations) is an
indication that it is likely that a trend is present, whereas a
value of ${\it \rho} \leq$ $[2 (n-1)]^{-1/2}$ indicates that it is
unlikely that a trend is present. Baart therefore used the following
auto-correlation cut-off tolerance
\begin{equation}
{\it \rho}_{c}=\rho(cut)=[2(n-1)]^{-1/2}
\end{equation}

While computing the Fourier parameters of all the light curve data 
selected for the present analysis we have taken care of the fact  that 
Baart's condition is satisfied. The optimal order of the fit for RRc, RRab,
FU Cepheids (OGLE), First Overtone (FO) Cepheids, Eclipsing binaries and Mira
variables are 3, 5, 12, 10, 4 and 4 respectively.  The longer period data for FU Cepheids from Martin et al. (1979) and Moffett et al. (1998)
are fitted with fifth order of the fit because of relatively small numbers of data points.
A typical example of the fitted light curves of all types of variables with the optimal order of the fit
is shown in Fig.~2.

All the data sets in Table 1 are finally fitted  with the optimal
order of the fit and the fitted light curves are used to derive the
Fourier phase and amplitude parameters from the Fourier
coefficients. Fig. 1 shows the fitted light
curves of FU Cepheids. Although the number of data points for
the longer period are less, the phase coverage is satisfactory to do the Fourier decomposition. Although the phase coverage is
poor, the fits are reasonably good. The lower right panel shows the
example of a short period fundamental mode Cepheid from the OGLE-III
database which has a good phase coverage. ${\chi_{\nu}^{2}}$ is the Chi Square per degree of freedom ($\nu$) of the fit. The degree of freedom ($\nu$) is the number of data points minus the number of parameters of the fit.
The Fourier decomposition parameters ($a_i, b_i$) for Cepheids have been computed based on the optimal
order of the fit by the calculation of the unit-lag auto-correlation
function
\begin{figure}
\centering
\includegraphics[height=9cm,width=9cm]{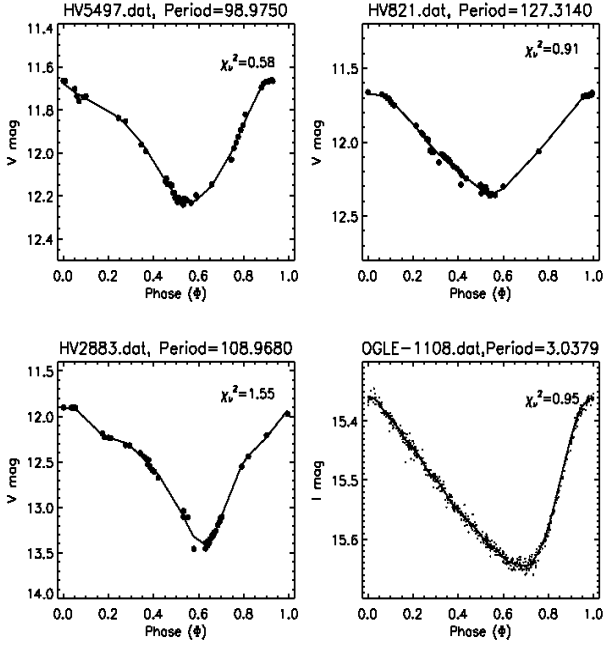}
\caption{Fitted light curves for fundamental mode long period
Cepheids from Moffett et al. (1998).}
 
\label{Fig 1 }
\end{figure}
\begin{figure}
\centering
\includegraphics[height=9cm,width=9cm]{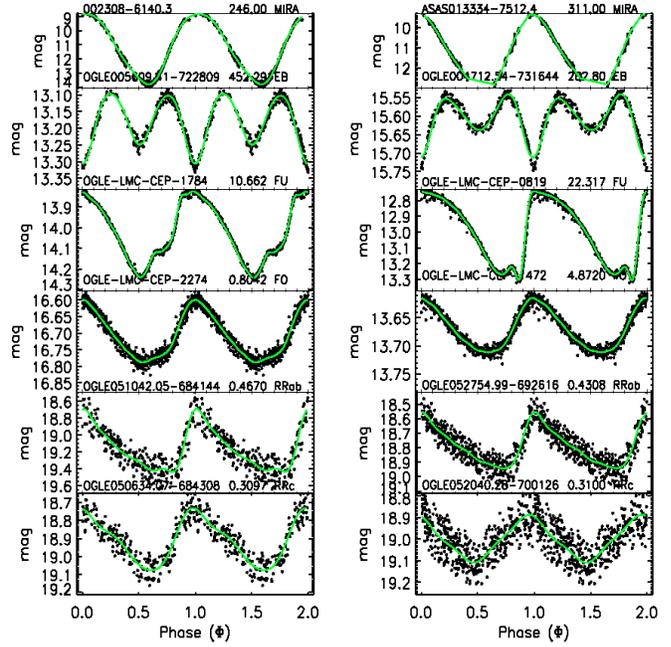}
\caption{Fitted light curves of different classes of variables used in the analysis
obtained with the optimal order of the fit. The caption at the top of each
panel shows the variable name, period and type of 
variables respectively. We have RR Lyrae variables (RRc, RRab), Cepheid
variables (Fundamental mode (FU) and First Overtone (FO)), Eclipsing binaries (EB)
and Mira variables (MIRA).
} 
\label{Fig 2 }
\end{figure}
\begin{figure}
\centering
\includegraphics[height=9cm,width=9cm]{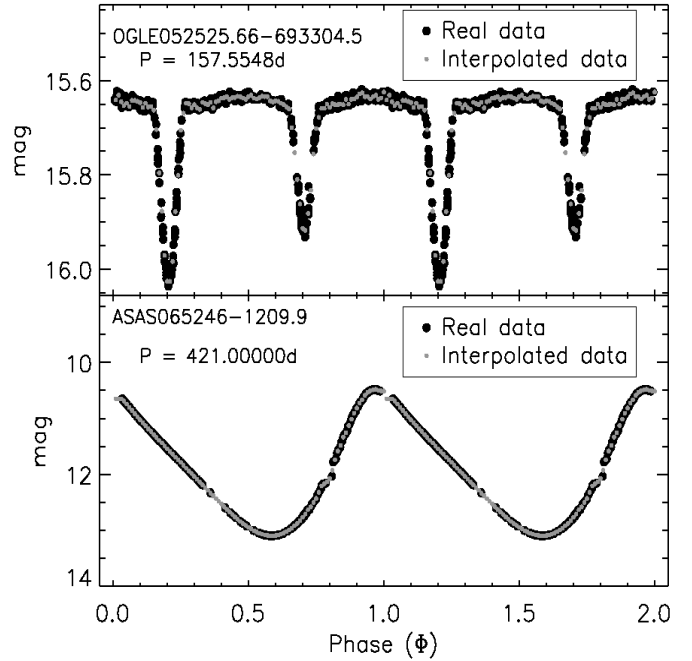}
\caption{Examples of interpolation of magnitudes for 100 points. The upper panel
shows the light curve with 100 interpolated data for the OGLE longer period
Eclipsing binary while the lower panel shows the interpolated data of a long
period Mira variable from the ASAS database. The 
lighter points denote the interpolated data while the bigger black dots 
represent the original data.}
\label{Fig 3 }
\end{figure}

\section{PRINCIPAL COMPONENT ANALYSIS}

The principal component analysis transforms the original set of $p$
variables by an orthogonal transformation to a new set of
uncorrelated variables or principal components (PCs). It involves a
simple rotation from the original axes to the new ones resulting in
principal components in decreasing order of importance. The first
few $q$ components ($q \ll p$) usually contain most of the variation in the
original data (Chatfield \& Collins 1980, Murtagh \& Heck 1987).
This feature of the PCA has been used in astronomical data analysis
primarily for the purpose of reducing the dimensionality of the data
and as a preprocessor for other automated
 techniques like Artificial Neural Networks (ANN). The application of PCA
 to the light curve analysis of variable stars has been limited to a few
 studies (Hendry et al. 1999, Kanbur et al. 2002, 2004, Tanvir et al. 2005).
 In the following, we briefly describe the transformation.

Let $m_{ij}$ be the $p$ magnitudes corresponding to $n$ light
curves. Let us define the $n \times p $ matrix
 by ${\textbf X} = x_{ij}$ ,
\begin{displaymath}
x_{ij}=\frac{m_{ij}-\overline{m}} {s_{
j}\sqrt{n}},
\end{displaymath}
with
\begin{displaymath}
\overline{m_{j}}= \frac{1}{n} \sum_{i=1}^{n} m_{ij},
\end{displaymath}
and
\begin{displaymath}
{s_{j}}^{2}=\frac{1}{n} {\sum_{i=1}^{n} {(
m_{ij}-\overline{m_{j}})}^{2}},
\end{displaymath}
where $\overline{m_{j}}$ is the mean value and $s_{j}$ is the
standard deviation. Using such standardization  we find the
principal components from the correlation matrix (cf. Murtagh \&
Heck 1987)
\begin{equation}
C_{jk}=\sum_{i=1}^{n} x_{ij} x_{
jk} =\frac{1}{n} \sum_{i=1}^{n} (m_{ij}-
\overline{m_{j}})(m_{ik}-\overline{m_{
k}})/( s_{j} s_{k}),
\end{equation}
 with the axis of maximum variance being the largest eigenvector $\textbf e_{1}$
 associated with the largest eigenvalue $\lambda_{1}$ of the equation
\begin{equation}
C \mathbf{e_{1}} = \lambda_{1} \mathbf{e_{1}}.
\end{equation}
The next (second) axis is to be orthogonal to the first and another
solution of  Eq. (8) gives the second largest eigenvalue
$\lambda_{2}$ and the corresponding eigenvector or the principal
component $\textbf e_{2}$. Hence the proportion of the total
variation accounted by the $j^{th} $ component is $\lambda_{j}/p$,
where $p$ is also the sum of the eigenvalues (Singh et al. 1998).

Let us suppose that the first $q$ principal components are
sufficient to retain the information on the original $p$ variables.
Therefore, we now have a ($p \times q$) matrix $\rm \mathbf{E_{\rm q}}$ of
eigenvectors. The projection vector $\textbf Z$ onto the $q$
principal components can be found by
\begin{equation}
\mathbf{Z}=\mathbf {x} \rm \mathbf{E}_{\rm q},
\end{equation}
where $\textbf x$ is vector of magnitudes defined by
\begin{displaymath}
x_{ij}\,{s_{j}\sqrt{n}} + \overline{m_{j}} = 
m_{ij},
\end{displaymath}
 and can be represented by
\begin{equation}
\mathbf x = \rm \mathbf {Z}{\rm \mathbf{E}_{\rm q}}^{\rm T}.
\end{equation}

We obtain the final light  curve  $\mathbf{x_{rec}}$ by multiplying
$x$ with $s_{j}\sqrt{n}$ and adding the mean. Z is a
($n \times q$) matrix and {$\rm \mathbf {E_{q}}$}$^{\rm\mathbf{T}}$ is a (
$q\times p$) matrix and hence the reconstructed light curve is
the original ($n\times p$) matrix. \\

With the phase ($\Phi$) as epoch for each light curve  available from Eq. (3), we interpolate and obtain $100$
 magnitudes for phase $0$ to $1$ in steps of $0.01$.
Therefore, each light curve now consists of $100$ data points (magnitudes)
normalized to unity. The input to the PCA are these 100 points of magnitudes
for each of the light curves.
We also emphasize that while applying PCA to the phased magnitudes of light 
curves, Fourier coefficients are not used to interpolate the light 
curves. We have used standard interpolation routines
in IDL for generating interpolated magnitudes in a light curve.
Two such examples of the result of interpolation are shown in Fig. 3.
The actual data points for the Mira 
variable (lower panel) are 223 while 100 interpolated magnitudes have been obtained.

\section{Analysis of light curves}
 In the subsequent analysis, we compare the capabilities of FD and PCA for structural analysis of Cepheids 
and classification accuracy for different classes of variable stars.
 
\subsection{Structural Analysis \& Classification}

\subsubsection{Fundamental mode (FU) Cepheids }
We use the light curve data for 1829 FU  classical
Cepheids  from various sources as mentioned in Table 1 (Data set
IIA+IIB+IIC). The majority of the data used in the analysis are from the OGLE
database. The Fourier decomposition of all the 1829  Cepheid light
curves has been independently done by us for the calculation of the
Fourier decomposition parameters as described in Sect. 2. We have seen
that all the Cepheid light curves selected in the present study give
satisfactory light curve shape with no numerical bumps or wiggles
when reconstructed using the Fourier parameters.

 PCA is performed on
an input matrix  consisting of a $1829 \times 100$ array
corresponding to 100 magnitudes from phase 0 to 1 for 1829
FU Cepheids. The result of the PCA output is shown in
Table 2. We see that first 10 PCs contain nearly 90
percent of the variance in the data. Fig. 4 shows the reconstruction
of four FU Cepheid light curves using the first 1, 3, 7 and 10 PCs.


\begin{table}
\caption{The first $10$ eigenvectors, their percentage of variance
and the cumulative percentage of variance of $1829$ fundamental mode
Cepheids. The input matrix is an $1829 \times 100$ array. }
\begin{tabular}{cccc}
\\
\hline \hline PC &Eigenvalue &Percentage& Cum. Percentage \\
\hline
1&41.0424&41.0424  &41.0424 \\
2&22.8331&22.8331  &63.8755 \\
3&11.7668&11.7668  &75.6423 \\
4& 5.4564& 5.4564  &81.0987 \\
5& 3.6225&  3.6225 &84.7212 \\
6& 2.4477&  2.4477 &87.1689 \\
7& 1.3398&  1.3398 &88.5087 \\
8& 0.7918&  0.7918 &89.3005 \\
9& 0.6435&  0.6435 &89.9440\\
10&0.6395& 0.6395 &90.5835 \\
\hline
\hline
\end{tabular}
\end{table}

\begin{figure}
\centering
\includegraphics[height=9cm,width=9cm]{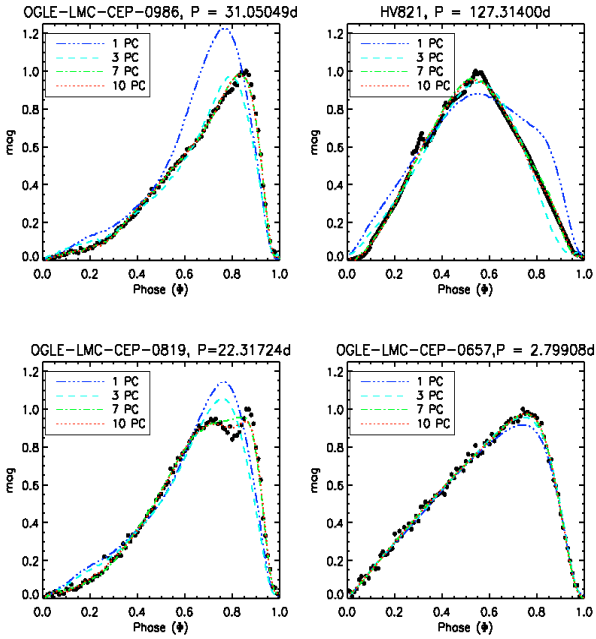}
\caption{Reconstruction of FU Cepheid light curves using the 
first 1, 3, 7 and 10 principal components. The input matrix is an array of
1829 rows (stars) and 100 columns (magnitudes from phase 0 to 1). The black
dots represent the original 100 interpolated data points normalized to unit 
magnitude.} 
\label{Fig 4}  
\end{figure}


\begin{figure}
\centering
\includegraphics[height=9cm,width=9cm]{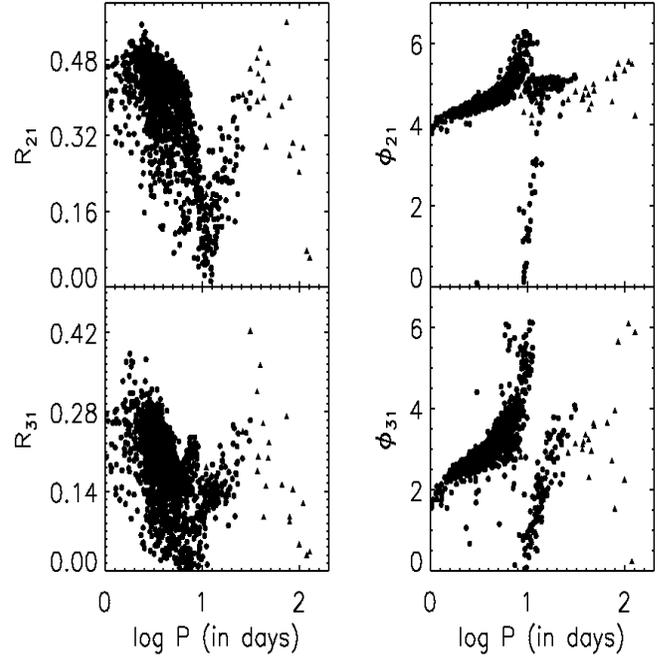}
\caption{Fourier parameters R$_{21}$, R$_{31}$, $\phi_{21}$,
$\phi_{31}$  as a function of log (Period) for the 1829 FU Cepheids (Data set IIA+IIB+IIC, Table 1). The Fourier parameters for the I band stars and V band stars are marked  with filled circles and filled upper triangles respectively.} 
\label{Fig 5}
\end{figure}
\begin{figure}
\centering
\includegraphics[height=9cm,width=9cm]{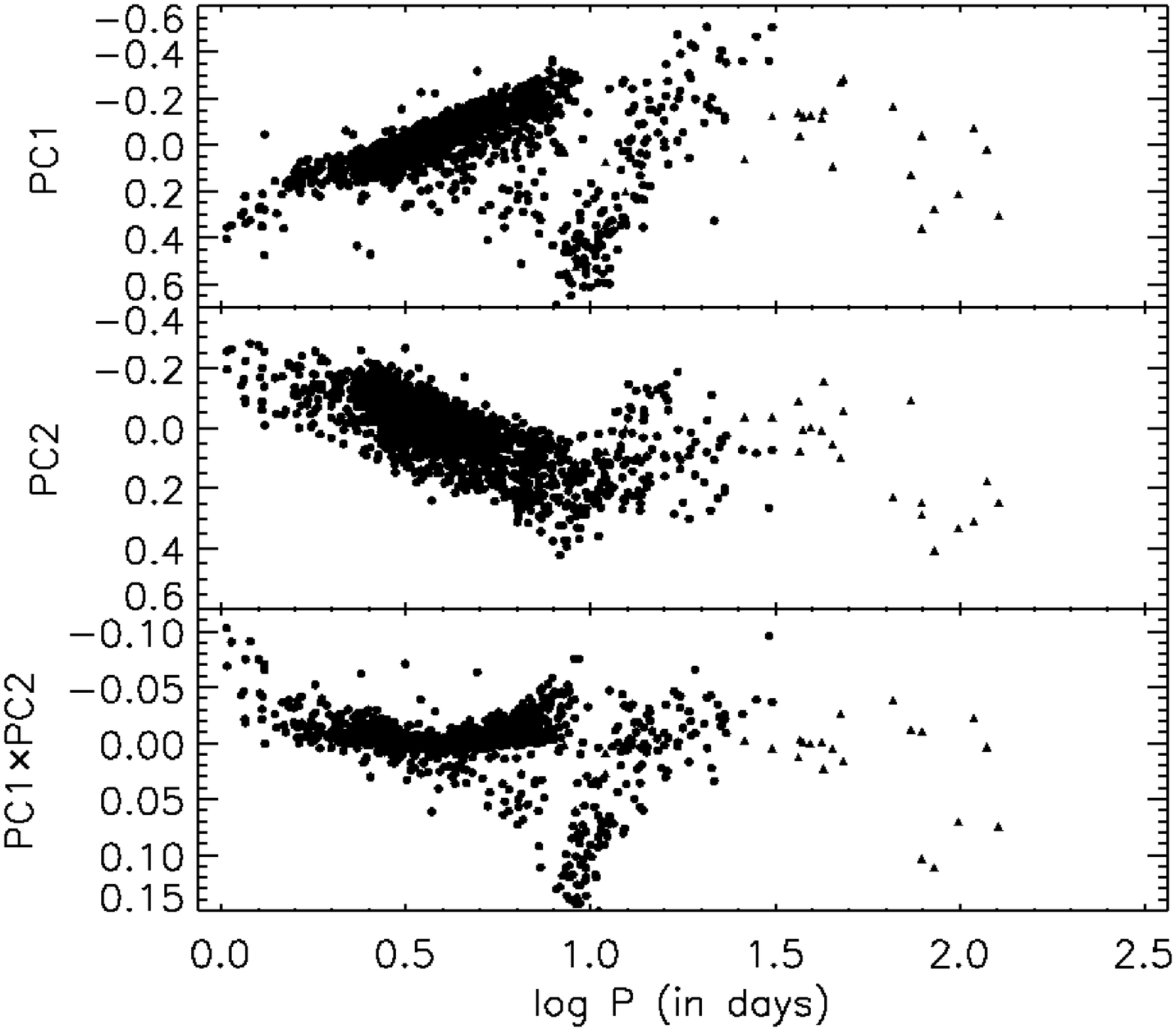}
\caption{First two PCs as a function of log (Period) for the 1829 FU Cepheids (Data set IIA+IIB+IIC, Table 1). The Fourier parameters for the I band stars and V band stars are marked  with filled circles and filled upper triangles respectively.}
 \label{Fig 6}
\end{figure}

Kanbur et al. (2002) have tried to explain the resonances using
the PCA on the Fourier coefficients {\bf($a_{i}, b_{i}$)}. But due to the relatively 
smaller number of data points they did not give any definite conclusions about 
some of the resonances suggested by  Antonello \& Morelli (1996) in the period 
range 1.38$\,< \log$~P \,$ < $ 1.43. By doing the PCA analysis of the same data as
used by Antonello \& Morelli (1996), Kanbur et al. (2002) could not
find any feature in that period range. Based on the available light curves
covering a wide range of periods, we have plotted $R_{21}$,
$R_{31}$, $\phi_{21}$, $\phi_{31}$ versus $\log$ P in Fig. 5. It is
very evident from the plots that there is a definite structural
change in the Fourier coefficients at periods $\log$ P $\sim  1.0$
and  $1.5$, the latter being close to the  period range 1.38$\,< \log$~P \,$ < $ 1.43
 suggested by Antonello \& Morelli (1996). We see that
the Fourier decomposition parameters $R_{21}$ and $R_{31}$ decrease
till $\log$ P $\sim 1.0$, increase thereafter till $\log$ P $\sim
1.5$ and after that $R_{21}$ and  $R_{31}$ fall gradually again till 
$\log P \sim 2.10$. Similarly in the $\phi_{21}$ and $\phi_{31}$ plane, we see a
sharp discontinuity around $\log$ P $\sim 1$. The sharp and the more prominent
discontinuity around  $\log$ P $\sim 1.0$ is reflected in both  $\phi_{21}$ and $\phi_{31}$ plots, whereas the other changes in the light curve structures around the period $\log$ P $\sim 1.5$  are visible in all the Fourier parameter plots.

In Fig. 6 we plot the first two PCs and PC1$\times$PC2 (PC1x2) against $\log$ P. 
For  PC1, PC2 and PC1x2, a discontinuity around $\log$ P = 1.0 is 
quite visible. PC1, PC2 and  PC1x2 clearly show a change around the 
period $\log$ P $\sim$ 1.5. But the 
discontinuity around  $\log$ P $\sim$ 1 as revealed by the Fourier parameters $\phi_{21}$ and 
$\phi_{31}$ in Fig.~5 is much more pronounced as compared to the PC plots.

Kanbur et al. (2002), using the PCA analysis on the 
Fourier coefficients, did not find any structure changes in the period range 
$1.38 < \log$ P $< 1.43$. Using PCA on a larger light curve  data set  we have found that in fact there are structural 
changes around $\log$ P $\sim$ 1 and 1.5 and hence there  may exist resonances 
around these periods.  While the resonance around the period $\log$ P $\sim$ 1 is well-known, the first two PCs 
and PC1x2 show a change in the light curve structure  around 
$\log$ P = 1.5. It is difficult to pinpoint the exact location of the change in structure
 because of fewer stars in the period around $\log$ P $\sim$ 1.5. Model calculations are 
necessary to confirm the existence of this resonance.  Further, Antonello \& Poretti 
(1996) also used a number of data points of the longer period side and found 
some evidence of a decrease of R$_{21}$ at longer periods around 
($\log$ P $\sim 2$). It is difficult to confirm the existence of such a 
resonance from either FD or PCA  although we see some change in trend in the first two PCs around 
this period. Therefore, although there are changes in the light curve structures around the periods
$\log$ P $\sim$ 1.5 and 2.10 days, one cannot confirm the existence of resonances around these 
periods.  Such information about these resonances are generally derived from the combined photometric, 
spectroscopic observations and radiative hydrodynamical model calculations (Kienzle et al. 1999).

\subsubsection{First overtone (FO) Cepheids}

The light curves of FO Cepheids  show a 
discontinuity in the Fourier phase parameters
$\phi_{21}$ and $\phi_{31}$ around a period of $\sim$ 3.2 
days. This is shown in Fig. 7 for the OGLE data (Data set III) of 1228 FO Cepheids. This feature was  interpreted as the signature
of 2:1 resonance between the first and fourth overtones (Antonello \& Poretti 
1986).  This feature was however  not reproduced in the 
hydrodynamical models  and in the Fourier parameters of highly accurate 
observational radial velocity curves of FO Cepheids (Kienzle et al. 1999). 
By means of hydrodynamical models for FO Cepheids, Kienzle et al. (1999) have  
shown that the 3.2 day is not the resonance, the true resonance is at around  4.5 d and 3.2 d is not a resonance. On the other hand Buchler et al. ( 1996) had suggested that  for a consistent picture on the evolutionary 
Mass-Luminosity relations, the FO Cepheid resonance should occur at P = 4.3 
days . Therefore, not all such structures in the photometric Fourier parameters  need to be connected to the resonances.

On the other hand,  by analyzing the Fourier coefficients of a large number of 
 FO LMC Cepheids in the OGLE III database, Soszy\'{n}ski et al.
(2008) found a  change in the photometric Fourier parameters 
 around a period of  $\sim$ 0.35d . The short-period discontinuity at 0.35d can be 
explained by  presence of the 2:1 resonance between the first and fifth 
overtones in stars with masses of about 2.5 $M_\odot$ (Dziembowski \& Smolec 
2009).

In Fig. 7, we plot the Fourier parameters R$_{21}$, $\phi_{21}$,
R$_{31}$, $\phi_{31}$ for 1228 LMC FO Cepheids (Data set
III in Table 1). The optimal order of the fit to the Fourier method
has been found to be 10. There is a definitive marked structure of
discontinuity in the Fourier plots at periods around 0.35 and 3.2
days.

 We now try to find out whether our PCA procedure can extract
the information about the structure changes. We carry out the PCA on a $1228
\times 100$ matrix of 1228 LMC FO Cepheids with 100 I band magnitudes
corresponding to phase 0 to 1 in steps  of 0.01. Fig. 8  shows the plot of 
 first three PCs versus the period.  A sharp 
discontinuity around the shorter period end near 0.35 day is evident in all
the PC plots. Also, some change in the light curve 
structure seems to occur near to 4 days for all the PC plots.  There is no  change in the 
light curve structure around 3.2 days in PC2 and PC3 whereas in PC1,  there is a change in the light curve shape around a period of $\sim$ 3.2 days

Thus, we see that the Fourier parameters performed better in bringing out the structural changes in FU Cepheids while for FO Cepheids  the performance of FD and PCA techniques is similar.

\begin{figure}
\centering
\includegraphics[height=9cm,width=9cm]{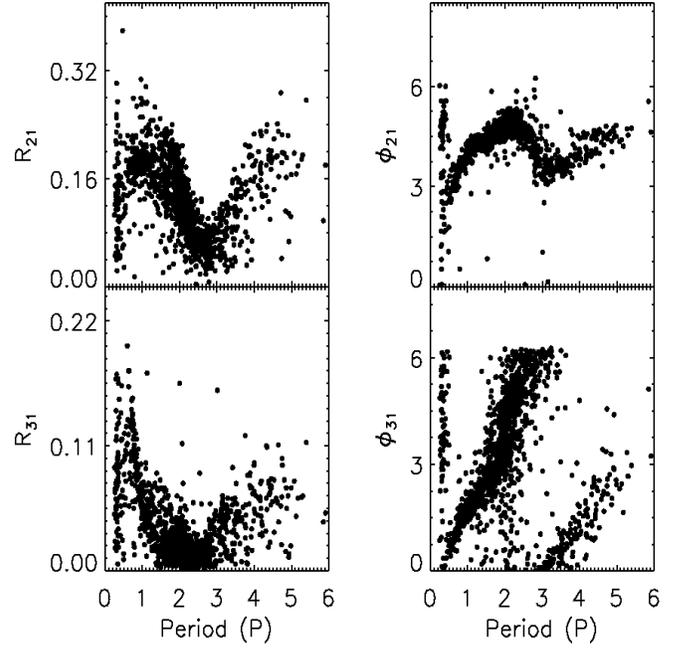}
\caption{Fourier parameters  R$_{21}$, R$_{31}$, $\phi_{21}$,
$\phi_{31}$  as a function of log (Period) for 1228  LMC  overtone
Cepheids (Data set III).}
\label{Fig 7 }
\end{figure}
\begin{figure}
\centering
\includegraphics[height=12cm,width=9cm]{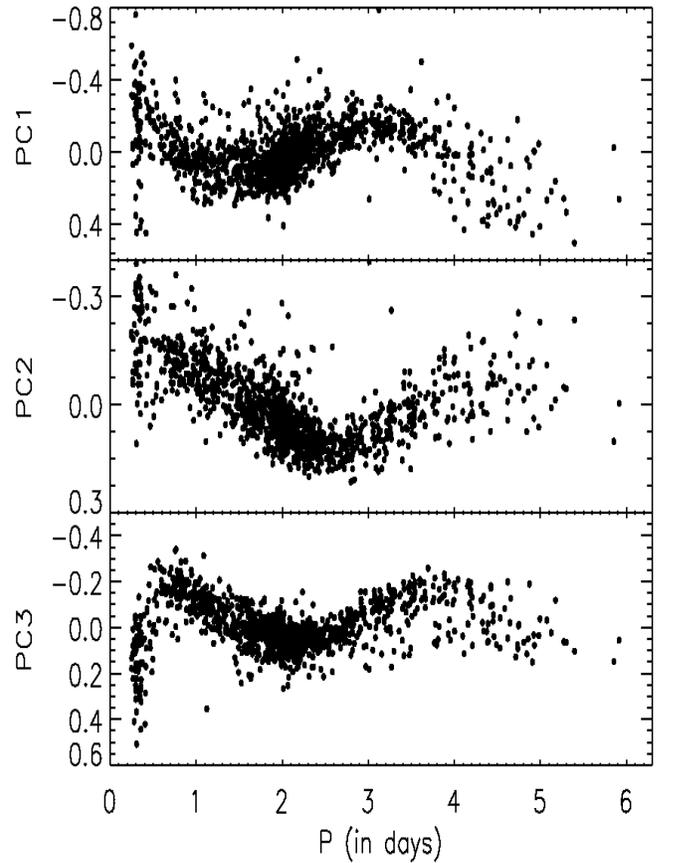}
\caption{First three PCs versus Period for LMC overtone Cepheids (Data
set III). The change in the light curve shape as shown  in Fig. 7
are also seen from the PC plots. The input matrix is an array of
1228 rows (stars) and 100 columns (magnitudes from phase 0 to 1).}
\label{Fig 8 }
\end{figure}
\subsubsection{Classification}
We now explore the possibility of classification of different
classes of variable stars on the basis of FD \& PCA.
We use the Fourier decomposition parameter R$_{21}$ and the first principal component PC1 to classify  all the 17,606 stars of different variability classes in Table 1. 
In Fig.~9 we plot the Fourier parameter R$_{21}$ versus $\log  $ P.  
As may be seen, the Mira variables form a separate group because of their longer periods and not because of separation in R$_{21}$.
However, in the intermediate period range (1 to 100 days), Eclipsing binaries have distinct R$_{21}$ values from other classes of variables.
Fig.~10  shows plot of $\log R_{21}$ to demonstrate the complete range of R$_{21}$ for 4085 Eclipsing binaries. In the short period range 
there is a considerable overlap between the FO Cepheids and RRc stars.  This degeneracy in the Fourier parameter R$_{21}$  in the short 
period range cannot be lifted and  the classification accuracy cannot be improved by further manipulation.

We carry out the  PCA on a 17606$\times$100 matrix of 17,606 stars, each star having 100 values of magnitudes in their light curves. 
We have used the first principal component (PC1) as it contains the maximum variance in this data set. As in the case of FD, the PCA is able to separate
 the Mira variables and the Eclipsing binaries and the separation is more effective in the case of PCA (Fig.~11).
The plot of PC1-$\log$ P space also  shows that although PC1 is able to separate the Eclipsing binaries and Mira variables, there is some 
overlap in the regions dominated by RR Lyraes and Cepheids. 
 In the next step, we choose only the samples of RR Lyraes (RRab \& RRc) and Cepheids (FU \& FO) that could not be 
separated well by the use of PCA on the whole data set. We now run PCA on 10,643 light curves (Data set  IA+IB+IIA+IIB+IIC+III) of RR Lyraes and Cepheids.
The result of PCA  on this 10643$\times$100 array  is shown in Fig.~12.  It may be noted that PC1 is able to separate FU Cepheids and RRab
stars to  a large extent while there is some overlap between RRc and FO Cepheids in a narrow period range (0.25-0.5 d). 
We hope to return to this degeneracy problem in a subsequent study in which we also intend to increase the sample by adding more classes of  
variables.  
\begin{figure}
\centering
\includegraphics[height=9cm,width=9cm]{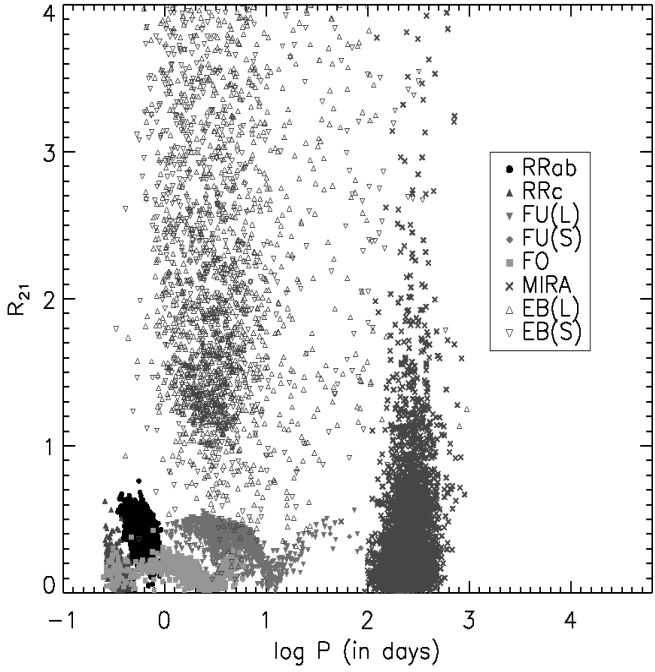}
\caption{The classification based on  R$_{21}$
obtained from the FD method. L and S denote the LMC and SMC objects 
respectively.}
\end{figure}
\begin{figure}
\centering
\includegraphics[height=9cm,width=9cm]{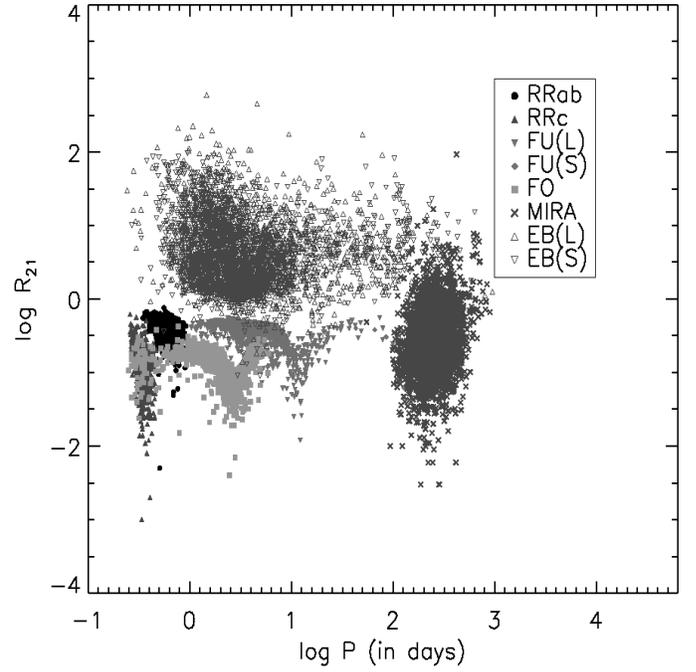}
\caption{The classification based on  $\log R_{21}$
obtained  from the FD method. }
\end{figure}



\begin{figure}
\centering
\includegraphics[height=9cm,width=9cm]{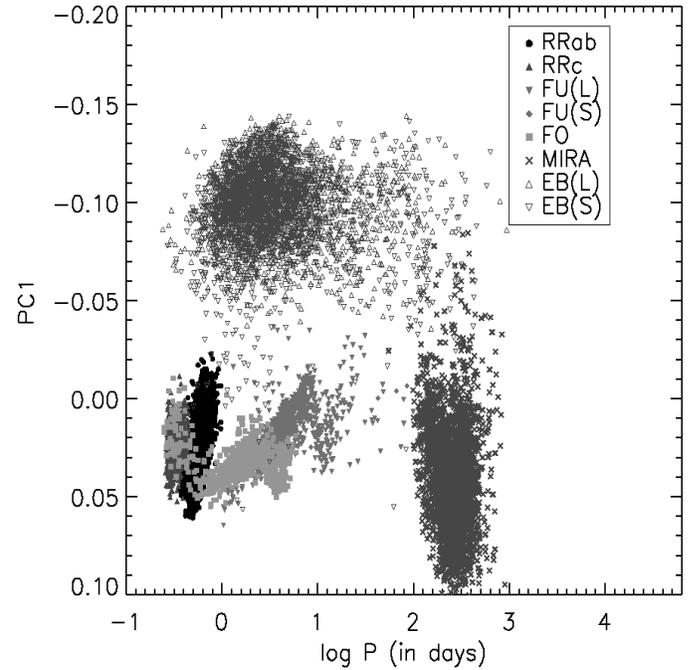}
\caption{The classification based on  PC1 obtained  from PCA of 
 100 interpolated magnitudes for the phase from 0 to 1 in
steps of 0.01.}
\end{figure}
\begin{figure}
\centering
\includegraphics[height=9cm,width=9cm]{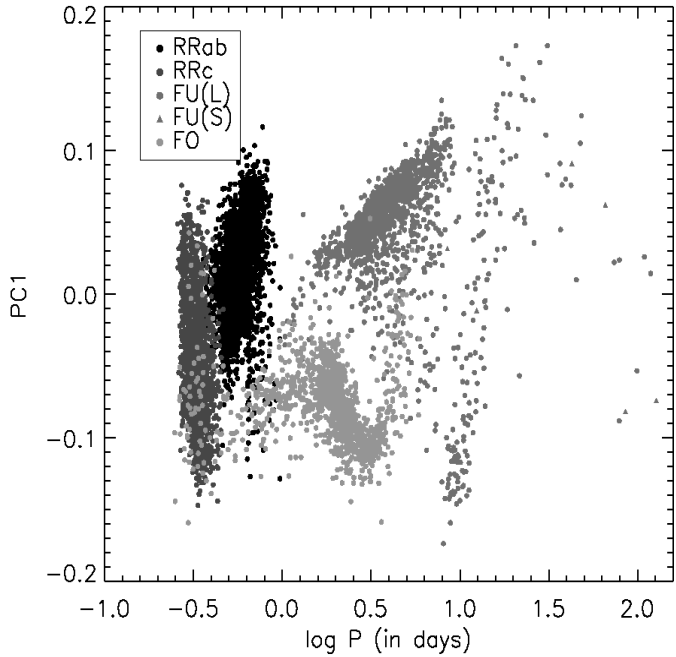}
\caption{The classification between RR Lyraes (RRab \& RRc) and Cepheids 
(FU \& FO) based on PC1. }
\end{figure}

\section{Conclusions}
Fourier decomposition is a trusted and much applied technique for
analyzing the behaviour of light curves of periodic variable stars.
It is well suited for studying individual light curves as the
Fourier parameters can be easily determined. However, when the
purpose is to tag a large number of stars for their variable class
using photometric data from large surveys, the technique becomes
slow and cumbersome and each light curve has to be fitted
individually  and then analyzed. The same is true if the aim is to look
for resonances in the light curves in an automated way for a large
class of pulsators. It is, therefore, desirable to look for methods
that are reliable, automated and unsupervised and can be applied to
the available light curve data directly.

Some attempts have been made in the recent past to use the well
known PCA for the light curve analysis, but
the major drawback of these studies was that they required the
calculation of the Fourier parameters which then went as input to
the PCA. This meant that the PCA, which was supposed to replace the
Fourier decomposition, in fact relied on it. Also for precise and accurate
determination of Fourier parameters,  the light curve should
have good phase coverage and less noisy data points so that the fit to the
light curve is good enough to rely on its parameters.  But this is not the case for each and every light curve data generated from the automated surveys.
Sometimes there are gaps and/or outliers in the data. The fitting of such a light curve will give a wrong estimation of the Fourier parameters.

In this paper we have used the original light curve data for computation of the  principal components. It involves four simple steps{\bf:} a) to phase every  
light  curve between 0 to 1 with respective period in days. b) Interpolation of light curve magnitudes in short steps (0.01) between phase 0 to 1  
to obtain 100 points of magnitude for each light curve. c) Normalize the magnitudes
between 0 and 1 for each of the light curves and  d) to do PCA on the
normalized magnitudes of 100 points for all the light curves.

The PCA is then used to analyse the structure of  the light curves  of classical Cepheids and the results compared with those obtained from the
analysis of the Fourier parameters. In addition, the  two techniques are compared with their ability to classify stars into different variability 
classes.

We  applied the PCA technique  to study the structure of light curves of
fundamental and first overtone 
Cepheids.  By choosing a large data set of a large range of periods
we have shown that the structure of the fundamental mode Cepheid
light curves shows significant changes around the  periods $\log P
\sim$ 1 and 1.5. The resonance around the period  $\log P \sim$ 1 is
well known. The first two PCs also show that the behavior of the
light curves changes around the period $\log P \sim$ 1.5 which is
close to the resonance suggested by Antonello \& Poretti (1996) in
the period range 1.38$\,< \log$~P \,$ < $ 1.43. There is some  evidence of
the structural change in the light curve shape around the period
$\log P \sim$ 2.0 also but this can be confirmed only when longer
period data become available. We find that  the Fourier parameters performed better in 
bringing out the discontinuities in FU light curves at period around $\log P \sim$ 1. 

For the first \, overtone LMC Cepheids, we find a discontinuity at a
shorter period of $\sim$ 0.35d. The first few PCs also show a clear
trend of structural changes of the first overtone Cepheids at this
short period.  For FO Cepheids, the performance of FD and PCA is similar in bringing out the structural changes around a period  of 0.35 day.
We have been able to find this  feature because of the
availability of significant number of light curves towards the
shorter period end of the LMC Cepheids in the OGLE database. 
The PCA technique can easily find similar resonances in the Galactic and SMC
first overtone Cepheids as and when there is substantial data
available for the short period objects of this class.

We have also demonstrated the ability of PCA and its distinct advantage over the FD method in classifying stars into different variability 
classes. Although alternative automated methods for variable stars classification exist, the PCA based technique can be used as a first step 
in hierarchical classification scheme because of its accuracy and efficiency. 

Data compression ratio using PCA on the direct light curve data is
enormous, a fact that has great relevance when dealing with large
databases of the future. Also, we have shown some preliminary
results of variable star classification for an ensemble of  17,606
stars selected in the present analysis. In a future  paper, we will describe
the application of the PCA technique with a larger, more diverse database by
looking at the classification accuracy and errors.

\section*{Acknowledgments}

The authors thank an anonymous referee for helpful comments
which improved the presentation of the paper and  Manabu Yuasa for 
correcting the first draft. Igor Soszy\'nski is thanked for providing 
valuable information regarding the OGLE data. SD thanks Council of Scientific 
\& Industrial Research (CSIR), Govt. of  India, for a Senior Research Fellowship. HPS thanks CRAL-Observatoire de Lyon for an invited Professorship.

\end{document}